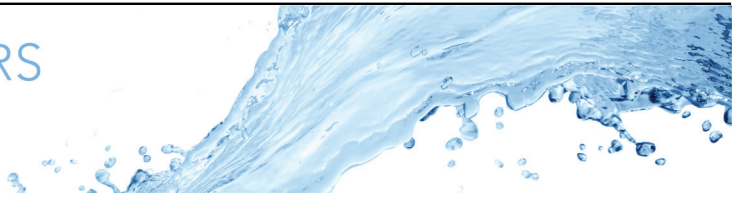

# Three-dimensional generative adversarial networks for turbulent flow estimation from wall measurements

**Antonio Cuéllar**[1],†, **Alejandro Güemes**[1], **Andrea Ianiro**[1], **Óscar Flores**[1], **Ricardo Vinuesa**[2] and **Stefano Discetti**[1]

[1]Department of Aerospace Engineering, Universidad Carlos III de Madrid, 28911 Leganés, Spain
[2]FLOW, Engineering Mechanics, KTH Royal Institute of Technology, 10044 Stockholm, Sweden

Different types of neural networks have been used to solve the flow sensing problem in turbulent flows, namely to estimate velocity in wall-parallel planes from wall measurements. Generative adversarial networks (GANs) are among the most promising methodologies, due to their more accurate estimations and better perceptual quality. This work tackles this flow sensing problem in the vicinity of the wall, addressing for the first time the reconstruction of the entire three-dimensional (3-D) field with a single network, i.e. a 3-D GAN. With this methodology, a single training and prediction process overcomes the limitation presented by the former approaches based on the independent estimation of wall-parallel planes. The network is capable of estimating the 3-D flow field with a level of error at each wall-normal distance comparable to that reported from wall-parallel plane estimations and at a lower training cost in terms of computational resources. The direct full 3-D reconstruction also unveils a direct interpretation in terms of coherent structures. It is shown that the accuracy of the network depends directly on the wall footprint of each individual turbulent structure. It is observed that wall-attached structures are predicted more accurately than wall-detached ones, especially at larger distances from the wall. Among wall-attached structures, smaller sweeps are reconstructed better than small ejections, while large ejections are reconstructed better than large sweeps as a consequence of their more intense footprint.

† Email address for correspondence: acuellar@ing.uc3m.es

## 1. Introduction

The ubiquitous nature of turbulent flows motivates the need for control to enhance the performance of a wide variety of devices. However, closed-loop control of turbulent flows (Choi, Moin & Kim 1994) requires continuous monitoring of their state. It is of utmost importance to be able to sense the flow state with minimal intrusiveness. Sometimes non-intrusive sensing is the only option available. This is the case of wall-bounded flows, making it possible to embed sensors within the wall. Non-intrusive sensing of turbulent flows has been the subject of several studies in the past decades. Machine learning has revolutionized the field of fluid mechanics (Brunton, Noack & Koumoutsakos 2020; Mendez *et al.* 2023), including both experiments (Discetti & Liu 2022; Vinuesa, Brunton & McKeon 2023) and simulations (Vinuesa & Brunton 2022). As such, the recent advances in machine learning and the wealth of available computational resources offer new interesting avenues for flow sensing.

The estimation of flow velocity solely on the basis of wall measurements was first explored using linear methods, such as linear stochastic estimation (LSE; Adrian 1996). The use of LSE was successful for the reconstruction of large-scale wall-attached eddies (Baars, Hutchins & Marusic 2016; Suzuki & Hasegawa 2017; Encinar, Lozano-Durán & Jiménez 2018; Illingworth, Monty & Marusic 2018; Encinar & Jiménez 2019). This methodology is capable of reconstructing a certain range of length scales of the structures populating the vicinity of the wall (i.e. the buffer layer) with reasonable accuracy. In the region farther from the wall, only large-scale motions are generally captured. These reconstructions can be more sophisticated by supplementing the methodology with further instruments in order to manipulate the filtering of scales, retaining and targeting the reconstruction over a broader spectrum. For example, in the work by Encinar & Jiménez (2019) with a turbulent channel flow in a large computational domain at a high friction Reynolds number, the large-scale structures containing about 50 % of the turbulent kinetic energy and tangential Reynolds stresses are reconstructed successfully up to $y/h \approx 0.2$, while only attached eddies of sizes of the order of $y$ are sensed in the logarithmic layer.

An alternative linear approach is the extended proper orthogonal decomposition (EPOD) (Borée 2003), which can establish a correlation between input and output quantities through the projection of their corresponding proper orthogonal decomposition (POD) modes (Lumley 1967). Despite leveraging only linear correlation, EPOD presents the advantage of being able to target specific significant features in a space of reduced dimensionality. A non-exhaustive list of examples of EPOD applications to the reconstruction of turbulent flows includes the estimation of the low-dimensional characteristics of a transonic jet (Tinney, Ukeiley & Glauser 2008), wakes behind wall-mounted objects (Bourgeois, Noack & Martinuzzi 2013; Hosseini, Martinuzzi & Noack 2016), wing wakes (Chen, Raiola & Discetti 2022), turbulent channel flows (Discetti, Raiola & Ianiro 2018; Güemes, Discetti & Ianiro 2019) and even high-Reynolds-number pipe flows (Discetti *et al.* 2019). The limitations in terms of reconstruction capabilities and spectrum range found with EPOD are similar to those with LSE.

Lasagna *et al.* (2015) studied multiple-time-delay estimation techniques. Although linear methods provide accurate estimations in the viscous layer, nonlinearities must be considered to extend the reconstruction into the buffer layer. Also, Chevalier *et al.* (2006) and Suzuki & Hasegawa (2017) highlighted the importance of incorporating nonlinear terms to get a more accurate estimation with a Kalman filter. Following the seminal work by Milano & Koumoutsakos (2002), neural networks emerge as an alternative able to cope with nonlinear relations between sensor and flow features. Recently, deep-learning algorithms have been leveraged for flow reconstruction from sensors. For example, the laminar wake of a cylinder and the flow in a turbulent channel have been reconstructed successfully in two dimensions from coarse measurements with convolutional neural networks (CNNs) (Fukami, Fukagata & Taira 2019, 2021). The performances of LSE and CNNs in estimation from wall measurements in a turbulent channel flow have been compared by Nakamura, Fukami & Fukagata (2022), reporting that linear models can provide comparable results at the cost of establishing a nonlinear framework to combine and provide the inputs to the system. Nevertheless, the use of nonlinearities through CNNs can be very effective and neural networks seem more robust against noise than LSE. Burst events in the near-wall region such as ejections and sweeps were studied by Jagodinski, Zhu & Verma (2023), with a three-dimensional (3-D) CNN capable of predicting their intensities, and also providing information about the dynamically critical phenomena without any prior knowledge. For the specific task of estimation of flow velocity from wall sensors, Güemes *et al.* (2019) proposed using CNNs to estimate temporal coefficients of the POD of velocity fields. This approach has shown to be superior to EPOD, achieving better accuracy at larger distances from the wall. Guastoni *et al.* (2021) compared the performances of estimators based on a fully convolutional network (FCN) to estimate the velocity fluctuations directly, or to estimate the field through POD modes (FCN-POD), using as input pressure and shear-stress fields at the wall. The FCN and FCN-POD have shown remarkable accuracy for wall distances up to 50 wall units at a friction-based Reynolds number $Re_\tau = 550$. Recently, Guastoni *et al.* (2022) explored this FCN architecture, but using the convective heat flux at the wall, reporting a 50 % error reduction at 30 wall units.

An additional improvement has been achieved by Güemes *et al.* (2021) with an algorithm based on generative adversarial networks (GANs; Goodfellow *et al.* 2014). This architecture consists of two agents, a generator and a discriminator, which are trained to generate data from a statistical distribution and to discriminate real from generated data, respectively. Generator and discriminator networks compete in a zero-sum game during the training process, i.e. the loss of one agent corresponds to the gain of the other, and vice versa. These GANs have been applied for variegated tasks in fluid mechanics in the last years, including super-resolution (Deng *et al.* 2019; Güemes, Sanmiguel Vila & Discetti 2022; Yu *et al.* 2022) and field predictions (Chen *et al.* 2020; Li *et al.* 2023).

In the work by Güemes *et al.* (2021), GANs are used to generate wall-parallel velocity fields from wall measurements – pressure and wall-shear stresses. This architecture has shown better performances than the FCN and FCN-POD architectures proposed earlier (Guastoni *et al.* 2021); furthermore, it has shown remarkable robustness in the presence of coarse wall measurements. This aspect is particularly relevant for the practical implementation in experimental and real applications where the spatial resolution of the sensors might be a limitation.

The main shortcoming of the aforementioned studies is that the estimation is carried out with planar data, i.e. the velocity is estimated on wall-parallel planes. An *ad hoc* network must thus be trained for each wall-normal distance. However, turbulent boundary layers are characterized by the presence of 3-D coherent features (Jiménez 2018), a fact that was first realized in the visual identification of the near-wall streaks by Kline *et al.* (1967). These structures follow a process of lift-up, oscillation and bursting, referred to as the near-wall energy cycle (Hamilton, Kim & Waleffe 1995), responsible for maintaining turbulence near the wall (Jiménez & Pinelli 1999). A similar cycle, albeit more complex and chaotic, was later identified in the logarithmic layer (Flores & Jiménez 2010), involving a streamwise velocity streak with a width proportional to its height that bursts quasi-periodically.

The search for organized motions and coherent structures in wall-bounded turbulent flows has resulted in several families of structures. The definition of many of these structures is based on instantaneous velocity fields, like the hairpin packets of Adrian (2007), or the more disorganized clusters of vortices of Del Álamo *et al.* (2006). Other structures, like the very large streaks of the logarithmic and outer region, have been described in terms of both two-point statistics (Hoyas & Jiménez 2006) and instantaneous visualizations (Hutchins & Marusic 2007). Of particular interest here are the Q-structures defined by Lozano-Durán, Flores & Jiménez (2012), which are based on a reinterpretation of the quadrant analysis of Willmarth & Lu (1972) and Lu & Willmarth (1973) to define the 3-D structures responsible of the turbulent transfer of momentum. These Q-structures are divided into wall-detached and wall-attached Qs events, in a sense similar to the attached eddies of Townsend (1961). As reported by Lozano-Durán *et al.* (2012), the detached Qs are background stress fluctuations, typically small and isotropic, without any net contribution to the mean stress. On the other hand, wall-attached Qs events are larger, and carry most of the mean Reynolds stress. Sweeps (Q4) and ejections (Q2) are the most common wall-attached Qs, appearing side by side in the logarithmic and outer regions.

It is reasonable to hypothesize that the nature of such coherent structures might have a relation with the capability of the GAN to reconstruct them or not. Employing state-of-the-art neural networks, the estimation of a full 3-D field from wall data requires the use of multiple networks targeting the reconstruction of wall-parallel planes at different wall distances. This implies bearing the computational cost of a cumbersome training of several networks, one for each of the desired wall-normal distances. Furthermore, each network is designed to reconstruct features at a certain distance from the wall, ignoring that the wall-shear stresses and pressure distributions depend also on scales located outside the target plane. The two-dimensional (2-D) reconstruction of an essentially 3-D problem complicates the interpretation of the actual scales that can be reconstructed in this process, while the continuity between adjoining layers – in terms of both absence of discontinuities within the field and mass conservation – is not necessarily preserved. Some recent works tackle similar problems, also from a 3-D perspective, such as the reconstruction of an unknown region of the flow through continuous assimilation technique by Wang & Zaki (2022), the reconstruction of fields from flow measurements by Yousif *et al.* (2023), and the reconstruction from surface measurements for free-surface flows through a CNN by Xuan & Shen (2023).

This work aims to overcome the aforementioned limitations by leveraging for the first time a full 3-D GAN architecture for 3-D velocity estimation from the wall. We employ a dataset of 3-D direct numerical simulations (DNS) of a channel flow. The reconstruction capabilities of a 3-D GAN are assessed. Section 2 describes both the training dataset and the 3-D GAN networks employed in the present study, while § 3 reports and discusses the results both in terms of reconstruction error statistics and in terms of structure-specific reconstruction quality. The physical interpretation of the results is given in terms of the framework of quadrant analysis in three dimensions (Lozano-Durán *et al.* 2012). Finally, § 4 presents the conclusions of the study.

## 2. Methodology

### 2.1. *Dataset description*

The dataset employed in this work consists of 3-D flow fields and shear and pressure fields at the wall of a minimal-flow-unit channel flow. Our numerical simulations are performed with a state-of-the-art pseudo-spectral code that uses a formulation based on the wall-normal vorticity and the Laplacian of the wall-normal velocity, and a semi-implicit Runge–Kutta for time integration (Vela-Martín *et al.* 2021). The solver uses a Fourier discretization in the wall-parallel directions and seventh-order compact finite differences in the wall-normal direction with spectral-like resolution (Lele 1992). The simulation domain is a periodic channel with two parallel walls located $2h$ apart, with sizes $\pi h$ and $\pi h/2$ in the streamwise and spanwise directions, respectively. This small channel fulfils the conditions established in the work by Jiménez & Moin (1991), which defines the minimal channel unit able to sustain turbulence.

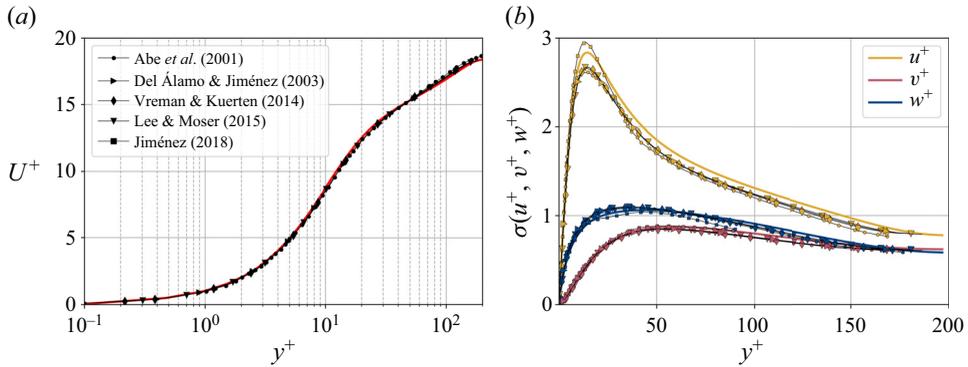

Figure 1. Wall-normal profiles of (*a*) the mean streamwise velocity and (*b*) standard deviation $\sigma$ of the three velocity components. Data are presented in inner units and compared to other databases at a similar $Re_\tau \approx 180$, including a minimal channel unit (Jiménez 2018) and several bigger channels.

In this work, we indicate with *x*, *y* and *z* the streamwise, wall-normal and spanwise directions, respectively, with their corresponding velocity fluctuations denoted by *u*, *v* and *w*. Simulations are performed at a friction-based Reynolds number $Re_\tau = u_\tau h/\nu \approx 200$, where $\nu$ refers to the kinematic viscosity, and $u_\tau = \sqrt{\tau_w/\rho}$ indicates the friction velocity, with $\tau_w$ the average wall-shear stress, and $\rho$ the working-fluid density. The superscript $+$ is used to express a quantity in wall units. To ensure statistical convergence and to minimize the correlation between the fields employed, data were sampled every $\Delta t^+ \approx 98$, i.e. 0.5 eddy-turnover times. The mean streamwise profile and the standard deviation of the velocity components are shown in figure 1. Moreover, the mean-squared velocity fluctuations in inner units are plotted in figure 6, where they can be compared with those reported for similar channel flows at $Re_\tau \approx 180$ (Abe, Kawamura & Matsuo 2001; Del Álamo & Jiménez 2003; Vreman & Kuerten 2014; Lee & Moser 2015).

Both the wall pressure $p_w$ and the wall-shear stress in the streamwise ($\tau_{w_x}$) and spanwise ($\tau_{w_z}$) directions are used for the flow field estimations. The data are fed into the proposed network on the same grid as the simulation. The streamwise and spanwise directions are discretized each with 64 equally spaced points, while the volume is discretized with 128 layers with variable spacing from the wall to the mid-plane. This discretization provides a set of grid points with a similar spacing to that found at $Re_\tau = 180$ in the work by Del Álamo & Jiménez (2003). The estimation capability of the 3-D fields was tested for volumes occupying the whole wall-parallel domain, but with different ranges in the wall-normal direction, giving rise to a set of test cases. These test cases are sketched in figure 2 and summarized in table 1, with $N_x$, $N_y$ and $N_z$ indicating the size of the mesh along each direction, and $\Delta y^+_{min}$ and $\Delta y^+_{max}$ being respectively the minimum and maximum wall-normal lengths of each grid step within the domain of each of the cases. Starting from the wall, cases A to C progressively reduce their wall-normal top limit from $y/h < 1$ to $y/h < 0.21$, so as the number of $x$–$z$ layers. Case D is defined as the domain complementary to that of case C, i.e. covering wall-normal distances in the range $0.21 < y/h < 1$.

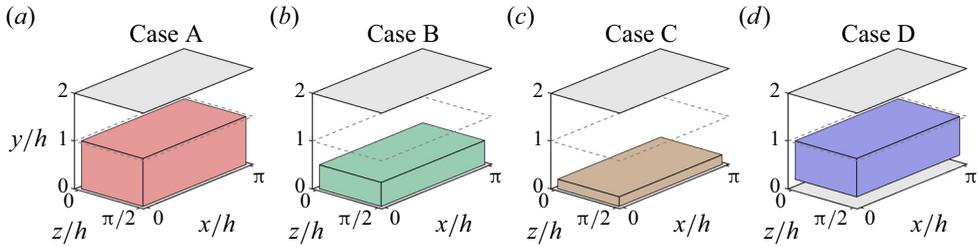

Figure 2. Representation of the reconstructed volume of the channel in each case, as defined in table 1.

| Case | $y/h$ range | $N_y$ | $N_x$ | $N_z$ | $\Delta y^+_{min}$ | $\Delta y^+_{max}$ |
|------|-------------|-------|-------|-------|--------------------|--------------------|
| A | 0–1 | 64 | 64 | 64 | 0.48 | 6.5 |
| B | 0–0.52 | 48 | 64 | 64 | 0.48 | 5.0 |
| C | 0–0.21 | 32 | 64 | 64 | 0.48 | 2.6 |
| D | 0.21–1 | 32 | 64 | 64 | 2.8 | 6.5 |

Table 1. Details about the domains of the cases, as represented in figure 2.

### 2.2. *Generative adversarial networks*

In this work, a GAN architecture is proposed to estimate 3-D velocity fields from wall measurements of pressure and shear stresses. The implementation details of the proposed architecture are presented below, being an extension to the 3-D space of the network proposed in the work of Güemes *et al.* (2021).

A schematic view of the generator network $G$ can be found in figure 3. The network is similar to that proposed by Güemes *et al.* (2021), although with some modifications. It is fed with wall measurements and consists of 16 residual blocks, containing convolutional layers with batch normalization layers and parametric-ReLU. The classic ReLU activation function provides as output $f(x) = x$ for positive entries and $f(x) = 0$ (flat) for negative ones. Parametric-ReLU does the same on the positive input values, while for negative entries it is defined as $f(x) = ax$, where $a$ is a parameter (He *et al.* 2015). In addition, sub-pixel convolution layers are used at the end for super-resolution purposes, adding more or fewer layers depending on the resolution of the fed data. To deal with 3-D data, a third spatial dimension has been added to the kernel of the convolutional layers. Since the present dataset does not require the network to increase the resolution from the wall to the flow in the wall-parallel directions, the sub-pixel convolutional layers present after the residual blocks in Güemes *et al.* (2021) have been removed. Similarly, the batch-normalization layers were dropped since they were found to substantially increase the computational cost without a direct impact on the accuracy (He *et al.* 2016; Kim, Lee & Lee 2016).

The increase of the wall-normal thickness up to the desired output volume has been achieved by using blocks composed of up-sampling layers followed by convolutional layers with parametric-ReLU activation, which we will refer to as up-sampling blocks throughout this document. For cases A, B and D, the first block is placed before the residual blocks.

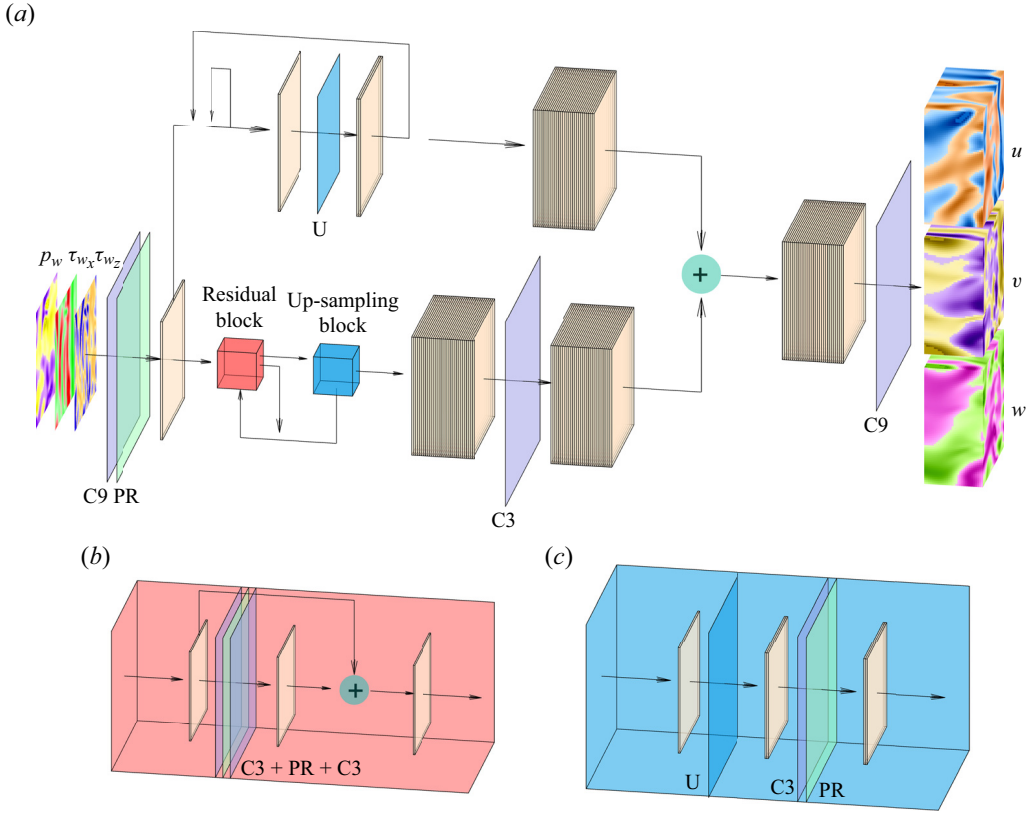

Figure 3. (*a*) Sketch of the generator network. (*b*) The residual block and (*c*) the up-sampling block sub-units, which are repeated recursively through network (*a*). The filter dimension is represented only in the network input $[p_w, \tau_{w_x}, \tau_{w_z}]$ and output $[u, v, w]$. All other layers work over 64 filters, except the last layer, which only has 3 filters coinciding with the output. The planar panels indicate the different layers of the network: up-sampling (U), parametric-ReLU (PR), and convolution layers with kernel sizes 3 (C3) and 9 (C9), respectively. Arrows indicate the flow of data through layers.

They increase the size of the domain by a factor of 2 in all cases except for the first up-sampling block in case B, which increases the size of the domain by a factor of 3. The rest of the up-sampling blocks are applied after the residual blocks, whose indexes are specified in table 2, together with the number of trainable parameters of the networks. The number of residual blocks, which has been increased to 32 with respect to Güemes *et al.* (2021), and the criterion to decide when to apply the up-sampling blocks, are analysed with a parametric study, for which a summary can be found in Appendix A.

A schematic of the discriminator network *D* is presented in figure 4. This network is very similar to that proposed by Güemes *et al.* (2021). The main difference is the change of the convolutional kernel to the 3-D space, including this new dimension. It consists mainly of a set of convolutional layers that progressively reduce the size of the domain and increase the number of filters. Then, with a flatten layer and two fully-connected layers, the network provides a single output in the range 0–1. Further details can be found in Appendix A. Additionally, it is important to note that due to the wall-normal dimension of its input data, one discriminator block was removed from cases C and D, which led to the counter-effect of increasing the number of trainable parameters reported in table 2. This network makes use of the leaky-ReLU activation function (Maas, Hannun & Ng 2013; He *et al.* 2015).

| | Residual block up-sampling scheme | Trainable parameters | |
|---|---|---|---|
| Case | | $G$ | $D$ |
| A | 0-6-12-18-24-30 | $9.0 \times 10^6$ | $18.2 \times 10^6$ |
| B | 0-5-10-15-20 | $8.0 \times 10^6$ | $18.2 \times 10^6$ |
| C | 6-12-18-24-30 | $8.0 \times 10^6$ | $23.8 \times 10^6$ |
| D | 0-7-14-21-28 | $8.0 \times 10^6$ | $23.8 \times 10^6$ |
| 2-D GAN | — | $7.3 \times 10^5$ | $8.0 \times 10^7$ |

Table 2. Details on the implementation of the architectures. The column 'Residual block up-sampling scheme' indicates the indexes of the residual blocks that are followed by an up-sampling block. The number of trainable parameters of the generator ($G$) and discriminator ($D$) networks are also reported. Cases A–D are compared with the 2-D GAN from Güemes *et al.* (2021).

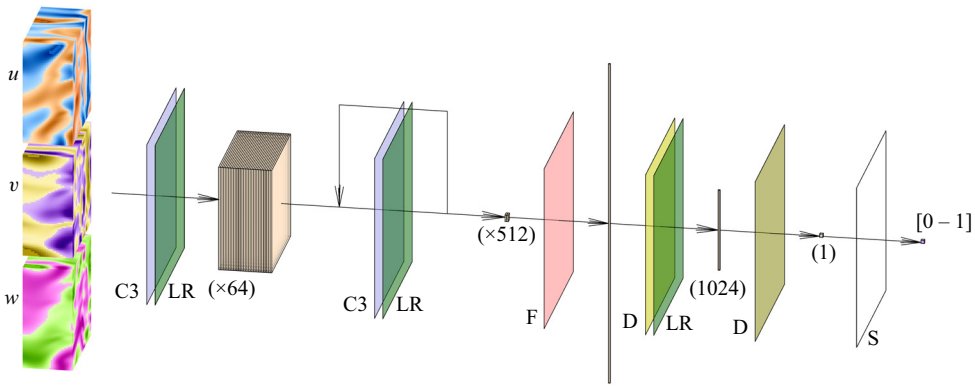

Figure 4. Sketch of the discriminator network. The network receives as input the velocity-fluctuation fields. The planar panels indicate the different layers of the network: the data passes through a set of convolutional (C3) and leaky-ReLU (LR) layers, reducing the dimension of the domain in the $x$, $y$ and $z$ coordinates as the number of filters increases progressively from 64 to 512. All these data are reshaped into a single vector with the flatten (F) layer. The dimensionality is reduced, first with a D fully connected layer with 1024 elements as output, and then with another D layer providing a single element, which is finally fed to a sigmoid (S) activation function.

The training process has been defined for 20 epochs, although the predictions are computed with the epoch where the validation loss stops decreasing, and the optimizer implements the Adam algorithm (Kingma & Ba 2014) with learning rate $10^{-4}$. In total, 24 000 samples have been used, keeping 4000 for validation and 4000 for testing. A random initial condition is set and evolved during about 100 eddy-turnover times to eliminate transient effects. Samples of the testing dataset are captured after approximately 100 eddy-turnover times from the last snapshot of the validation dataset to minimize correlation with the training data.

As mentioned above, the networks operate with the velocity fluctuations [$u$, $v$, $w$]. As there are significant differences in the mean values of the velocity components at the centre of the channel and in the vicinity of the wall, the mean values used to compute the field of fluctuation velocities have been obtained for each particular wall-normal distance $y$. In addition, to facilitate the training of the network, each fluctuating velocity component has been normalized with its standard deviation at each wall-normal layer (see figure 1). Similarly, the wall measurements $[p_w, \tau_{w_x}, \tau_{w_z}]$ provided to the network have been normalized with their mean value and standard deviation.

The training loss functions are defined as follows. The fluctuations of the velocity field can be represented as $\boldsymbol{u} = [u, v, w]$, such that $\boldsymbol{u}_{DNS}$ is the original field, and $\boldsymbol{u}_{GAN}$ is the field reconstructed by the generator network given its corresponding set of inputs $[p_w, \tau_{w_x}, \tau_{w_z}]$. With these two definitions, using the normalized velocity fields, the content loss based on the mean-squared error (MSE) is expressed as

$$\mathcal{L}_{MSE} = \frac{1}{3N_x N_y N_z} \sum_{i=1}^{N_x} \sum_{j=1}^{N_y} \sum_{k=1}^{N_z} \Big[ (u_{DNS}(i,j,k) - u_{GAN}(i,j,k))^2$$

$$+ (v_{DNS}(i,j,k) - v_{GAN}(i,j,k))^2 + (w_{DNS}(i,j,k) - w_{GAN}(i,j,k))^2 \Big], \quad (2.1)$$

Using the binary cross-entropy, an adversarial loss is defined as

$$\mathcal{L}_{adv} = -\mathbb{E}[\log D(\boldsymbol{u}_{GAN})], \quad (2.2)$$

to quantify the ability of the generator to mislead the discriminator, with $\mathbb{E}$ the mathematical expectation operator, and $D(\cdot)$ the output of the discriminator network when it receives a flow field as input – in this case, a GAN-generated flow field. This adversarial loss is combined with the content loss to establish the loss function of the generator network as

$$\mathcal{L}_G = \mathcal{L}_{MSE} + 10^{-3} \mathcal{L}_{adv}. \quad (2.3)$$

The loss function for the discriminator network, defined as

$$\mathcal{L}_D = -\mathbb{E}[\log D(\boldsymbol{u}_{DNS})] - \mathbb{E}[\log(1 - D(\boldsymbol{u}_{GAN}))], \quad (2.4)$$

also uses the binary cross-entropy to represent its ability to label correctly the real and generated fields. To ensure stability during the training process, both the adversarial and discriminator losses are perturbed by subtracting a random noise in the range 0–0.2. This technique, referred to as label smoothing, makes it possible to reduce the vulnerability of the GAN by modifying the ideal targets of the loss functions (Salimans *et al.* 2016).

## 3. Results

### 3.1. *Reconstruction accuracy*

The reconstruction accuracy is assessed in terms of the MSE of the prediction. The contribution of each velocity component ($\boldsymbol{u} = [u, v, w]$) to the metric presented in (2.1) has been computed along the wall-normal direction, denoted as $\mathcal{L}_u$, $\mathcal{L}_v$ and $\mathcal{L}_w$, respectively. In this case, the error is computed using only one component at a time, and the factor of 3 in the denominator is eliminated. As discussed in § 2.2, the training data have been normalized with their standard deviation for each wall-normal distance. This procedure allows us a straightforward comparison with the results of 2-D GAN architectures (Güemes *et al.* 2021). It must be remarked that flow reconstruction by Güemes *et al.* (2021) is based on an open-channel simulation; nonetheless, the similar values of $Re_\tau$ numbers provide a quite accurate reference.

The MSE for each velocity component is plotted with respect to the wall-normal distance for the selected network architectures of each case A–D in figure 5, and numerical data of the error at the wall distances used in the 2-D approach are collected in table 3 for comparison. They have been computed for the velocity fluctuations normalized with their standard deviation at each wall-normal coordinate $y^+$, allowing us to compare the accuracy of this network with the analogous 2-D study. Some general comments can be raised at first sight.

| $y^+$ | Case | $\mathcal{L}_u$ | $\mathcal{L}_v$ | $\mathcal{L}_w$ |
|---|---|---|---|---|
| 15 | A | 0.043 | 0.076 | 0.088 |
| | B | 0.038 | 0.066 | 0.074 |
| | C | 0.027 | 0.044 | 0.050 |
| | 2-D GAN | 0.013 | 0.018 | 0.019 |
| 30 | A | 0.137 | 0.214 | 0.205 |
| | B | 0.118 | 0.190 | 0.179 |
| | C | 0.095 | 0.152 | 0.143 |
| | 2-D GAN | 0.061 | 0.097 | 0.084 |
| 50 | A | 0.306 | 0.440 | 0.429 |
| | B | 0.277 | 0.411 | 0.395 |
| | D | 0.356 | 0.505 | 0.494 |
| | 2-D GAN | 0.185 | 0.289 | 0.268 |
| 100 | A | 0.639 | 0.788 | 0.782 |
| | B | 0.619 | 0.779 | 0.771 |
| | D | 0.665 | 0.815 | 0.813 |
| | 2-D GAN | 0.524 | 0.684 | 0.687 |

Table 3. The MSE for the three velocity components and for each case, at different wall distances. The results are compared with the 2-D analysis by Güemes *et al.* (2021). These quantities correspond to velocity fluctuations normalized with their standard deviation at each wall-normal coordinate $y^+$.

(i) As expected, and also reported in the 2-D analysis (Güemes *et al.* 2021), the regions closer to the wall show a lower $\mathcal{L}$. This result is not surprising: at small wall distances the velocity fields show high correlation with the wall-shear and pressure distributions, thus simplifying the estimation task for the GAN, independently on the architecture.

(ii) The streamwise velocity fluctuation $u$ always reports a slightly lower $\mathcal{L}$ than $v$ and $w$ for all the tested cases. This is due to the stronger correlation of the streamwise velocity field with the streamwise wall-shear stress.

(iii) The 3-D GAN provides a slightly higher $\mathcal{L}$ than the 2-D case. This was foreseeable: the 3-D architecture proposed here is establishing a mapping to a full 3-D domain, with only a slight increase in the number of parameters in the generator with respect to the 2-D architecture, as can be seen in table 2. Furthermore, there is a considerable reduction in the number of trainable parameters in the discriminator. If we consider $\mathcal{L}_u = 0.2$, then the reconstructed region with an error below this threshold is reduced from approximately 50 to slightly less than 40 wall units when switching from a 2-D to a 3-D GAN architecture.

In test case A, part of the effort in training is directed to estimating structures located far from the wall, thus reducing the accuracy of the estimation. For this reason, cases B and C were proposed to check whether reducing the wall-normal extension of the reconstructed domain would increase the accuracy of the network. Comparing the $\mathcal{L}$ values of cases A, B and C in table 3, it is observed that there is some progressive improvement with these volume reductions, although it is only marginal. For example, the error $\mathcal{L}_u$ at $y^+ = 100$ is

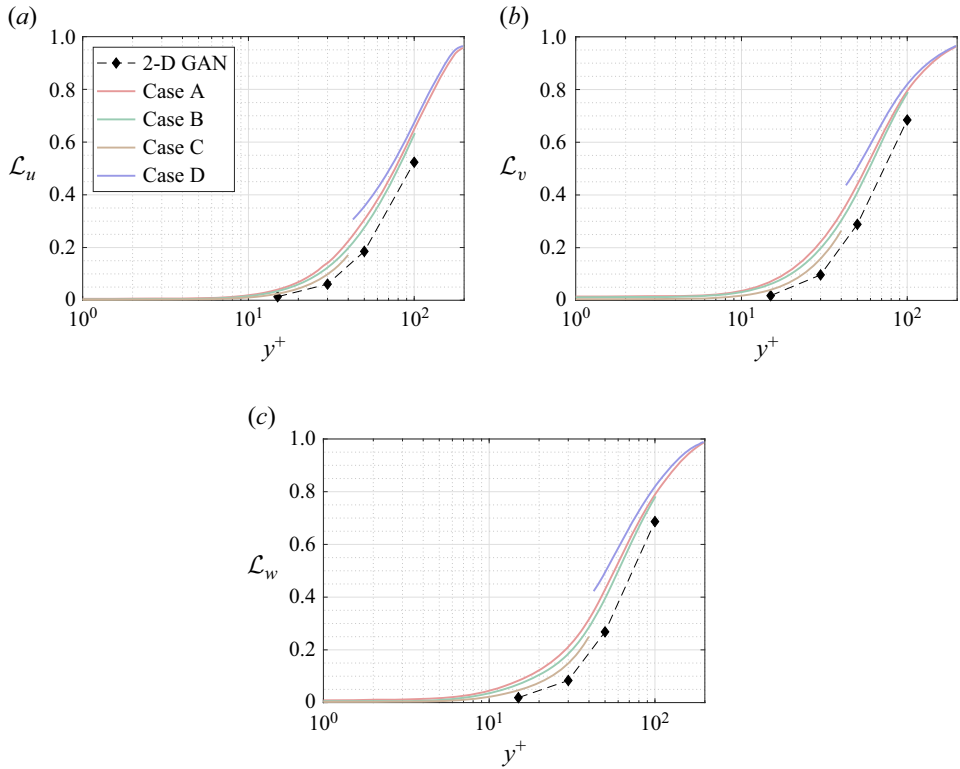

Figure 5. The MSE of the fluctuation velocity components (*a*) *u*, (*b*) *v* and (*c*) *w* for the 3-D GAN (continuous lines) and the 2-D GAN at $Re_\tau = 180$ (symbols with dashed lines) as implemented by Güemes *et al.* (2021). Velocity fluctuations are normalized with their standard deviation at each wall-normal coordinate $y^+$.

2 % lower when switching from case A to case B, i.e. reducing by a factor of 2 the size of the volume to be estimated. This fact can also be observed in figure 5(*a*), where the $\mathcal{L}_u$ values for all the cases can be compared directly. Similar conclusions can be drawn from the other velocity components. The improvement between cases is marginal. The quality of the reconstruction of one region seems thus to be minimally affected by the inclusion of other regions within the volume to be estimated. This suggests that the quality of the reconstruction is driven mainly by the existence of a footprint of the flow in a certain region of the channel. In Appendix B, we have included a comparison with LSE, EPOD and a deep neural network that replicates the generator of case A and provides an estimation of the effect of the discriminator. The accuracy improvement of the 3-D GAN with respect to the LSE and the EPOD is substantial, while the effect of the adversarial training does not seem to be very significant. Nevertheless, previous works with 2-D estimations (Güemes *et al.* 2021) have shown that the superiority of the adversarial training is more significant if input data with poorer resolution are fed to the network. We hypothesize that a similar scenario might occur also for the 3-D estimation; nonetheless, exploring this aspect falls outside the scope of this work.

Case D, targeting only the outer region, is included to understand the effect of excluding the layers having a higher correlation with wall quantities from the reconstruction process. The main hypothesis is that during training, the filters of the convolutional kernels may focus on filtering small-scale features that populate the near-wall region. Comparing the plots for cases A and D in figure 5, it is found that the $\mathcal{L}$ level in case D is even higher than in case A. While this might be surprising at first glance, a reason for this may reside in the difficulty of establishing the mapping from the large scale in the outer region to the footprint at the wall when such footprint is overwhelmingly populated by the imprint of near-wall small-scale features. Convolution kernels stride all along the domain, and when the flow field contains wall-attached events with a higher correlation, the performance far from the wall seems to be slightly enhanced. In case A, the estimator is able to establish a mapping for such small-scale features to 3-D structures, while for case D, such information, being uncorrelated with the 3-D flow features in the reconstruction target domain, is seen as random noise. This result reveals the importance of the wall footprint of the flow on the reconstruction accuracy.

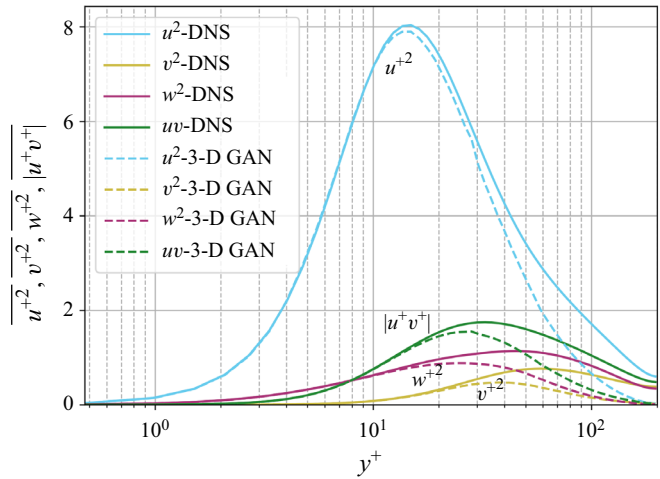

Figure 6. Mean-squared velocity fluctuations and shear-stress, given in wall-inner units.

Moreover, figure 6 shows the evolution of the mean-squared velocity fluctuations and the $u^+v^+$ shear stress of the reconstructed (case A) velocity fields with the wall-normal distance. As expected, far from the wall, the attenuation becomes more significant, while the accuracy is reasonable up to approximately 30 wall units, where the losses with respect to the DNS are equal to 4.0 % for $u^2$, 9.7 % for $v^2$, 12.8 % for $w^2$, and 6.9 % for $|uv|$. It is important to remark that the network is not trained to reproduce these quantities, as the loss function is based on the MSE and the adversarial loss. The losses reported in these quantities also may explain the MSE trends in figure 5, where the error grows with the wall-normal distance as the network generates more attenuated velocity fields. The fact that the kernels in the convolutional layers progressively stride along the domain implies that although the continuity equation might not be imposed as a penalty to the training of the network, the 3-D methodology exhibits an advantage with respect to the 2-D estimation. To assess this point, we compared the divergence of the flow fields obtained with the 3-D GAN with the divergence obtained from the velocity derivatives of three neighbouring planes estimated with 2-D GANs following Güemes *et al.* (2021). The standard deviation of the divergence, computed at both $y^+ = 20$ and $y^+ = 70$, is approximately 6 times smaller when employing the 3-D GAN.

An additional assessment of the results is made by comparing the instantaneous flow fields obtained from the predictions with those from the DNS. As an example, figure 7 shows 2-D planes of $\boldsymbol{u}$ at three different wall-normal distances, of an individual snapshot, according to case C. This case is selected for this example as it exhibits the best performance. In this test case, the attenuation of the velocity fluctuations is not significant. Up to the distances contained within case C, it is indeed possible to establish accurate correlations. On the contrary, the attenuation of the velocity field close to the centre of the channel is quite high (see figure 6). At $y^+ = 10$, it is difficult to find significant differences between the original and the reconstructed fields, with the smallest details of these patterns also being present. Farther away from the wall, at $y^+ = 20$, the estimation of the network is still very good, although small differences start to arise. At $y^+ = 40$, the large-scale turbulent patterns are well preserved, but the small ones are filtered or strongly attenuated.

In general, it can be observed that, regardless of the wall-normal location, the GAN estimator is able to represent well structures elongated in the streamwise direction (i.e. near-wall streaks), likely due to their stronger imprint at the wall. On the other hand, the $u$ fields at $y^+ = 40$ are also populated by smaller structures that do not seem to extend to planes at smaller wall-normal distances, thus indicating that these structures are detached. From a qualitative inspection, the detached structures suffer stronger filtering in the reconstruction process. Analogous considerations can be drawn from observation of the $v$ and $w$ components.

### 3.2. *Coherent structure reconstruction procedure*

Further insight into the relation between reconstruction accuracy and features of the coherent structures is provided by observing isosurfaces of the product $uv$ (the so-called uvsters; Lozano-Durán *et al.* 2012) reported in figure 8. Again, a sample of case C has been selected to compare the original structures (figure 8*a*) with the reconstructed ones (figure 8*b*). In both representations, structures of different sizes are observed, mainly aligned with the flow. The larger structures appear to be qualitatively well represented, while some of the smaller ones are filtered out or not well reproduced. Furthermore, it can be observed that the structures located farther from the wall are more intensively attenuated in the reconstruction process. Moreover, the majority of the structures and the volume identified within a structure are either sweeps or ejections.

The same procedure is followed for an instantaneous field reconstructed under case A, leading to figures 8(*c*) and 8(*d*), respectively. Similar phenomena are observed with some remarks. With a substantially larger volume, many more structures populate the original field. However, the reconstructed structures mainly appear close to the wall. Detached structures are filtered, while attached ones can be partially truncated in their regions farther away from the wall. Nevertheless, this instantaneous field also reveals that the filtering does not seem to be a simple function of the wall distance. Some of the attached structures recovered in the reconstructed field extend up far from the wall, and even up to the middle of the channel or beyond. This would not be possible if all types of structures were expected to be reconstructed up to a similar extent at any wall distance. The region far from the wall is depopulated using the same threshold for the isosurfaces, as the magnitude of the fluctuation velocities is strongly attenuated in this region.

For a quantitative assessment of the relation between coherent-structure features and reconstruction accuracy, here we follow an approach similar to that used by Lozano-Durán *et al.* (2012), based on a 3-D extension of the quadrant analysis (Willmarth & Lu 1972; Lu & Willmarth 1973), where turbulent structures are classified according to the quadrants defined in figure 9. A binary matrix on the same grid of the domain is built. The matrix contains ones in those points corresponding to a spatial position located inside a coherent structure, and zeros otherwise. Grid points where fluctuation velocities meet the following condition are within a structure

$$|\tau(x, y, z)| > H u'(y) v'(y), \quad (3.1)$$

where $\tau(x, y, z) = -u(x, y, z) v(x, y, z)$, the prime superscripts (′) indicate root-mean-squared quantities, and $H$ is the hyperbolic hole size, selected to be equal to 1.75. This is the same structure-identification threshold as in Lozano-Durán *et al.* (2012), for which sweeps and ejections were reported to fill only 8 % of the volume of their channel, although these structures contained around 60 % of the total Reynolds stress at all wall-normal distances. Without any sign criterion, this condition is used for the identification of all types of structures. Moreover, the signs of $u(y)$ and $v(y)$ are to be considered to make a quantitative distinction between sweeps (Q4) and ejections (Q2) as essential multi-scale objects of the turbulent cascade model that produce turbulent energy and transfer momentum.

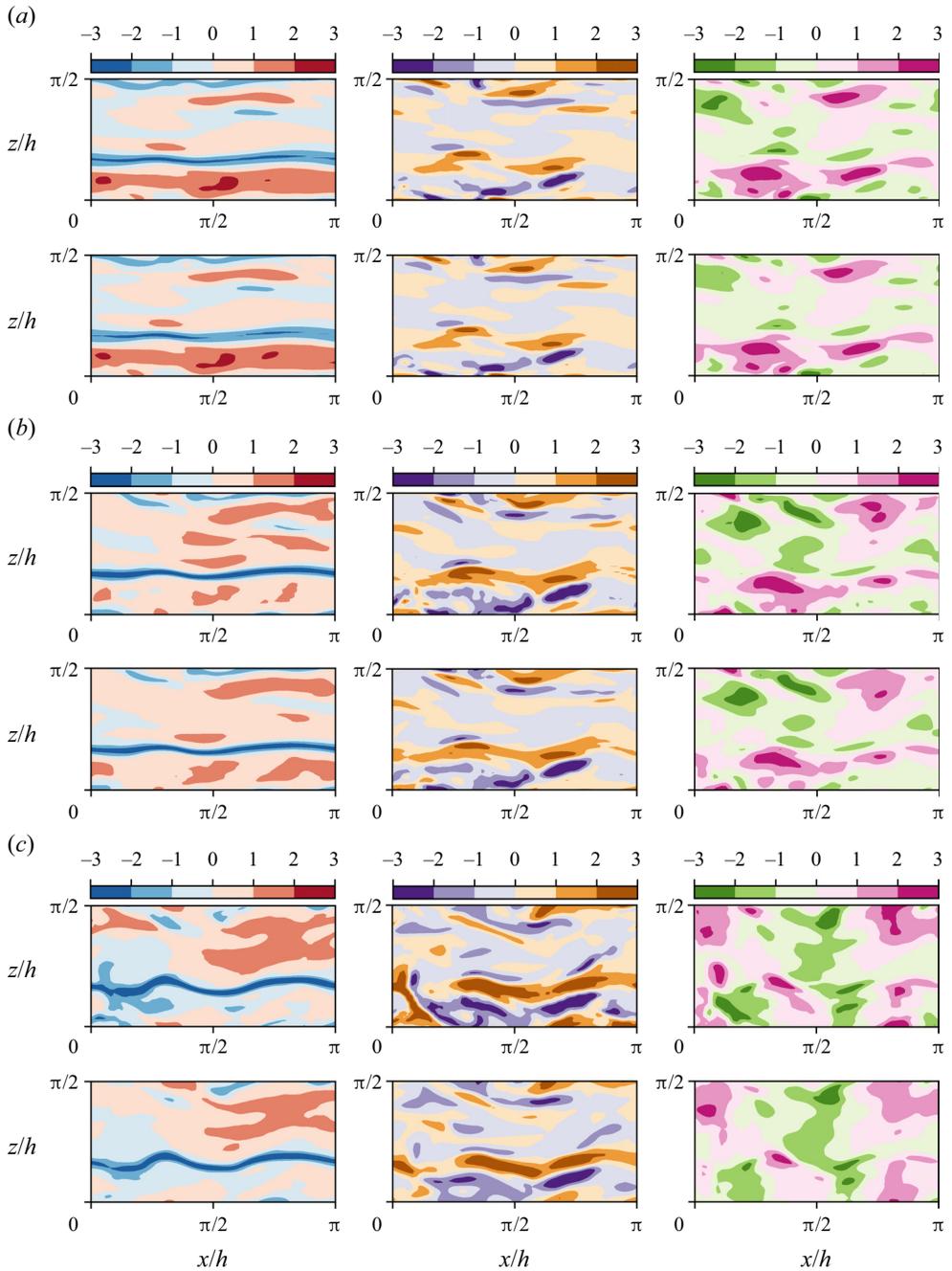

Figure 7. Instantaneous velocity fluctuations: $u$ (left), $v$ (centre) and $w$ (right). From each pair of rows, the top row is the original field from the DNS, and the bottom row is the field reconstructed with the GAN, for case C. Different pairs of rows represent 2-D planes at different wall-normal distances, with (*a*) $y^+ = 10$, (*b*) $y^+ = 20$ and (*c*) $y^+ = 40$. Instantaneous values beyond $\pm 3\sigma$ are saturated for flow visualization purposes. Velocity fluctuations are normalized with their standard deviation at each wall-normal coordinate $y^+$.

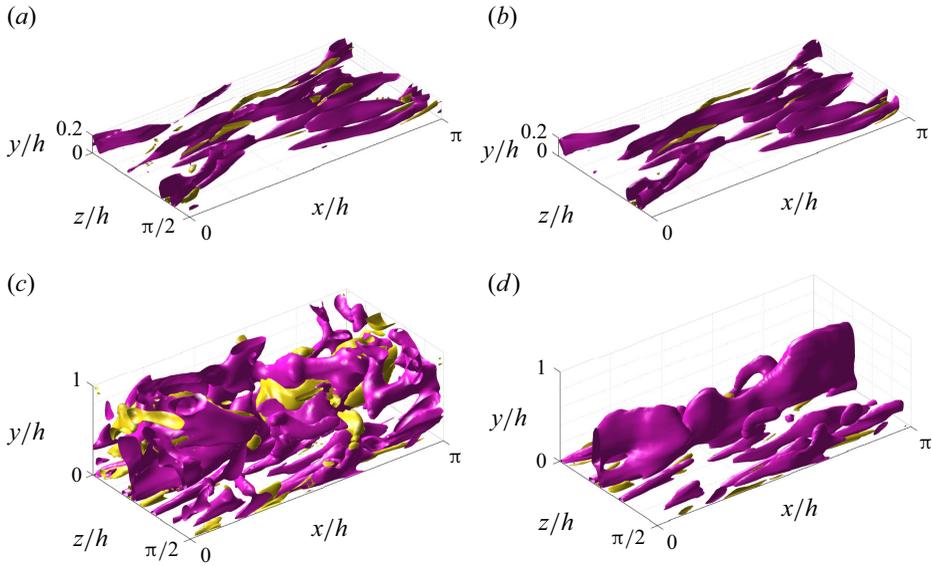

Figure 8. Instantaneous 3-D representation of the $uv$ field, for (*a*) original and (*b*) prediction from case C, and (*c*) original and (*d*) prediction from case A. Isosurfaces correspond to the 1.5 and $-1.5$ levels of $uv$, respectively, in yellow for quadrants Q1 and Q3, and in pink for Q2 and Q4.

The cells activated by (3.1) are gathered into structures through a connectivity procedure. Two cells are considered to be within the same structure if they share a face, a side or a vertex (26 orthogonal neighbours), or if they are indirectly connected to other cells. Some of the structures are fragmented by the sides of the periodic domain. To account for this issue, a replica of the domain based on periodicity has been enforced on all sides. To avoid repetitions, only those structures whose centroid remains within the original domain are considered for the statistics. Besides, structures with volume smaller than $10^{-5}h^3$ have been removed from our collection, to concentrate the statistical analysis on structures with a significant volume. Further conditions have been set to remove other small structures that are not necessarily included in the previous condition. Structures that are so small that they occupy only one cell – regardless of their position along $y$ and the cell size $\Delta y^+$ – and those whose centroid falls within the first wall-normal cell have been removed. Finally, structures that are contained within a bounding box of a size of one cell in any of the directions have been removed as well. For example, a structure comprising two adjoint cells delimited within a bounding box with two-cell size in one of the directions, but only one-cell size in the other two directions, would be discarded for the statistics. These restrictions still keep small-scale structures in the database, but eliminate those that are contaminated by the resolution errors of the simulation.

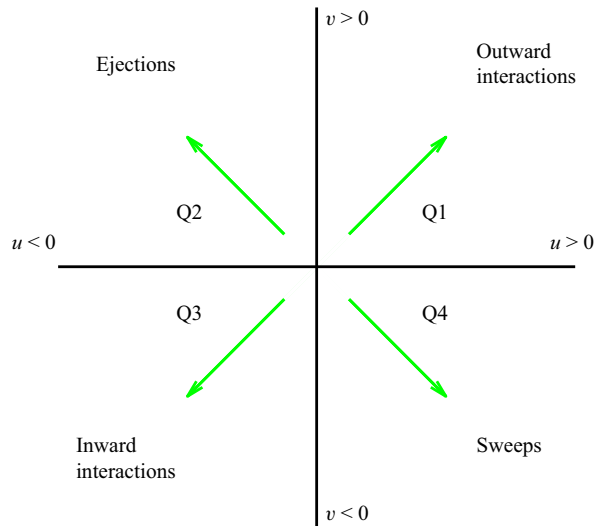

Figure 9. Quadrant map with the categorization of the turbulent motions as Qs events.

A statistical analysis on test case A has been carried out considering the different quadrants (see table 4). The figures reported by Lozano-Durán *et al.* (2012) can be used as a reference, although it must be remarked that discrepancies arise due to the differences in the fluid properties, the Reynolds number or the extension of the volume from the wall considered. The volume of a whole domain of case A is $4.93h^3$. The structures identified with (3.1) and the hyperbolic hole size used as threshold occupy only a small fraction of it, although they contain the most energetic part, able to develop and sustain turbulence. Note that the criterion established with this equation makes Qs (first row in table 4) not to be explicitly the sum of all individual Q1s, Q2s, Q3s, Q4s events – when no sign criterion is being applied over $u'$ and $v'$, without any distinction among different type of events. As seen with the instantaneous snapshots in figure 8, these statistics from 4000 samples tell us that Q1s and Q3s (see figure 9) are less numerous than negative Qs, both in volume and in units – these structures account for just 2 % in volume and 7 % in units in the work by Lozano-Durán *et al.* (2012). In addition, this analysis reveals that most of the volume fulfilling (3.1) belongs to Q2 structures, with Q4s occupying significantly less volume, while in unit terms the population of Q4s is approximately 50 % higher than that of Q2s. Individual Q2s, although fewer in number, are much bigger than Q4s. Structures have been considered as attached if $y_{min}/h \leqslant 0.1$ (figures 10a,d,g), where $y_{min}$ refers to the location of the closest point to the wall within a structure, and $y_{max}$ to the farthest one. All types of structures are notably attached in more than 60 % of the cases, with Q3s the most prone structures to be detached, and Q1s to be attached. Table 4 also offers a comparison between the target data from the DNS and the reconstruction from the 3-D GAN. There are no large discrepancies between target and prediction, with all the comments already mentioned applying to both of them. However, statistics are better preserved for Q1s and Q3s than for Q2s and Q4s. As expected, the GAN tends to generate slightly fewer Q2 and Q4 structures, a difference due mainly to wall-detached structures that are not predicted. However, these generated structures are bigger and occupy a larger volume than the original ones. As discussed below, not all the volume in the predicted structures is contained within the original ones.

| | | Volume of structures | | Number of structures | |
|---|---|---|---|---|---|
| | | Absolute [$h^3$] | Relative % | Total | Attached % |
| Qs | Target | 0.70 | 14.3 % | 14.5 | 68.9 % |
| | Prediction | 0.80 | 15.9 % | 13.3 | 77.4 % |
| Q1s | Target | 0.01 | 0.3 % | 3.85 | 90.3 % |
| | Prediction | 0.01 | 0.3 % | 3.80 | 90.7 % |
| Q2s | Target | 0.46 | 9.3 % | 5.24 | 81.0 % |
| | Prediction | 0.58 | 11.8 % | 4.59 | 89.4 % |
| Q3s | Target | 0.04 | 0.8 % | 3.10 | 68.4 % |
| | Prediction | 0.01 | 0.2 % | 3.12 | 69.5 % |
| Q4s | Target | 0.10 | 2.0 % | 7.41 | 81.5 % |
| | Prediction | 0.16 | 3.2 % | 7.11 | 85.2 % |

Table 4. Information about the structures identified with (3.1), Qs without any sign criterion on $u$ and $v$, and Q1s–Q4s for each quadrant. The DNS original data and the 3-D GAN prediction (case A) are compared with some statistics over the 4000 testing snapshots, considering the average volume occupied by structures per snapshot, their proportion of volume over the domain, the average number of structures in each snapshot, and the proportion of structures that are attached.

### 3.3. *Statistical analysis methodology*

A statistical analysis of the reconstruction fidelity of flow structures is carried out, with the previous condition (3.1) applied to the 4000 samples outside the training set. The structures found in the DNS- and GAN-generated domain pairs have been compared and matched. For each $i$th (or $j$th) structure, it is possible to compute its true volume $v_{T,i}$ (or predicted volume $v_{P,i}$) as

$$v_{T_i} = \sum_{x,y,z} \boldsymbol{T}_i \circ \boldsymbol{V}, \tag{3.2}$$

$$v_{P_j} = \sum_{x,y,z} \boldsymbol{P}_j \circ \boldsymbol{V}, \tag{3.3}$$

where the matrices $\boldsymbol{T}_i$ and $\boldsymbol{P}_i$ represent the target (DNS) and the prediction (GAN) domains for the $i$th structure, respectively, containing ones where the structure is present, and zeros elsewhere. Here, $\boldsymbol{V}$ is a matrix of the same dimensions containing the volume assigned to each cell. In a similar way, combining these two previous expressions, the overlap volume of two structures $i$ and $j$ within the true and predicted fields, respectively, is

$$v_{T_i,P_j} = \sum_{x,y,z} \boldsymbol{T}_i \circ \boldsymbol{P}_j \circ \boldsymbol{V}. \tag{3.4}$$

It must be considered that during the reconstruction process, the connectivity of regions is not necessarily preserved. This gives rise to a portfolio of possible scenarios. For instance, an original structure could be split into two or more structures in the reconstruction; small structures, on the other hand, could be merged in the estimated flow fields. Moreover, the threshold in (3.1) is based on the reconstructed velocity fluctuations, thus it can be lower than in the original fields. Hence all possible contributions from different structures overlapping with a single structure from the other dataset are gathered as follows:

$$\hat{v}_{T_i,P} = \frac{\sum_j v_{T_i,P_j}}{v_{T_i}}, \tag{3.5}$$

$$\hat{v}_{T,P_j} = \frac{\sum_i v_{T_i,P_j}}{v_{P_j}}. \tag{3.6}$$

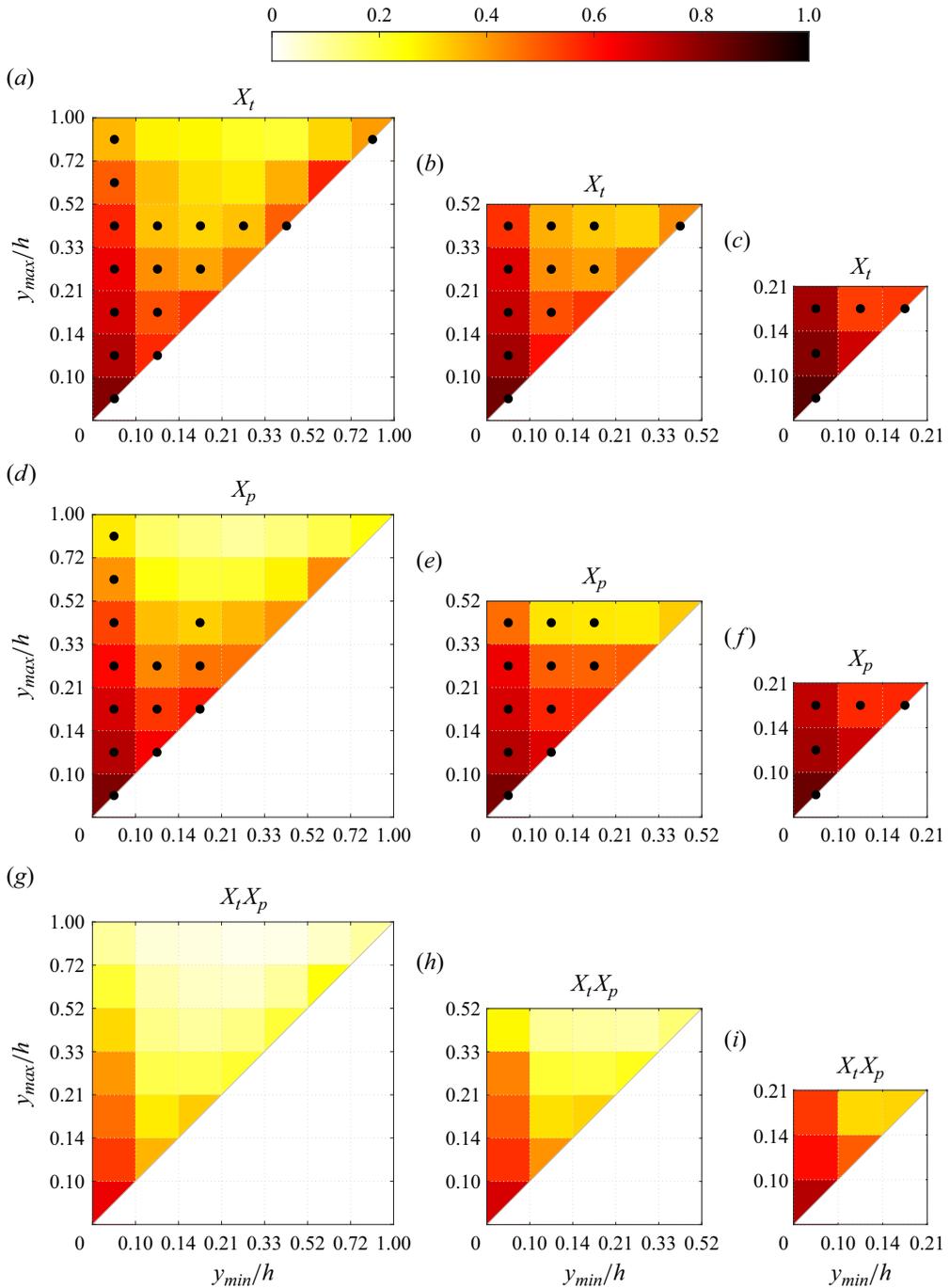

Figure 10. Maps of average density per bin of the matching quantities for cases (*a*,*d*,*g*) A, (*b*,*e*,*h*) B and (*c*,*f*,*i*) C, for the different metrics proposed. Dotted bins represent the top 95 % of the joint p.d.f. of the structures over that target or prediction set, respectively. Note that the scales in both axes are not uniform.

With the hat used to indicate the ratio, these metrics give the overlapped volume proportion of each structure $i$ from the target set, or $j$ from the prediction set, and are defined in such a way as some structures either split or coalesce. These structures are classified into intervals according to their domain in the $y$ direction, bounded by $y_{min}$ and $y_{max}$. Given the matching proportion of all the structures of each target-DNS and prediction-GAN set falling in each interval $(a, b)$ of $y_{min}$ and each interval $(c, d)$ of $y_{max}$, according to their individual bounds (respectively, $y_{min,i}$ and $y_{max,i}$, or $y_{min,j}$ and $y_{max,j}$), their average matching proportions $X_t$ and $X_p$ are computed:

$$X_{t,(a,b-c,d)} = \overline{\hat{v}_{T_i,P}} \quad \forall i \text{ such that } a < y_{min,i}/h < b, c < y_{max,i}/h < d, \tag{3.7}$$

$$X_{p,(a,b-c,d)} = \overline{\hat{v}_{T,P_j}} \quad \forall j \text{ such that } a < y_{min,j}/h < b, c < y_{max,j}/h < d. \tag{3.8}$$

Additionally, out of all these categories onto which the structures are classified according to their minimum and maximum heights, those contained within the top 95 % of the joint probability density function ( joint p.d.f.) have been identified with black dots in figures 10 and 11 to characterize the predominant structures in the flow.

### 3.4. *Analysis of the joint p.d.f.s of reconstructed structures*

The interpretation of the quantities defined in the previous subsection is as follows: $X_t$ is the proportion of the volume of the structures from the target set represented within the reconstructed structures; $X_p$ is the proportion of the volume of reconstructed structures matching the original ones. These quantities $X_t$ and $X_p$ are represented in figure 10 for each categorized bin and for cases A, B and C.

In figures 10($a,b,c$) (for $X_t$), the joint p.d.f. is compiled for the target structures, and in figures 10($d,e,f$) (for $X_p$), the joint p.d.f. is compiled for the predicted structures. The joint p.d.f. for the target structures indicates that the family of wall-attached structures (i.e. $y_{min} \leqslant 0.1h$) dominates the population, while wall-detached structures with $y_{min} \geqslant 0.1h$ do not extend far from the wall. Overall, the joint p.d.f.s are qualitatively similar to those obtained by Lozano-Durán *et al.* (2012) at higher Reynolds numbers, except for the wall-detached structures, which in the latter have $y_{max} - y_{min}$ approximately independent of $y_{min}$. This difference could be explained as an effect of the low friction Reynolds number, especially since the detached structures in Lozano-Durán *et al.* (2012) are linked to the dissipation process in the logarithmic and outer regions. Note that the bin highlighted at the top-right corner for the original set of structures is likely linked not to detached structures, but to tall attached structures rising from the opposite wall and extending beyond the middle of the channel.

Compared to the joint p.d.f. of the predicted set, the wall-detached structures of the original set extend farther from the wall. This suggests that the 3-D GAN may be losing the farthest region from the wall of some of the reconstructed structures, consistent with the flow visualizations presented in figure 8. The best reconstruction is reported in all cases for the shortest wall-attached structures. The values of $X_t$ and $X_p$ for wall-attached structures reduce progressively as $y_{max}/h$ increases. This trend is repeated for other columns of bins with $y_{min}/h \geqslant 0.10$, although the metrics are lower than for the wall-attached structures. The wall-detached structures that are contained within the top 95 % of the joint p.d.f. are reconstructed with modest values of $X_t$ and $X_p$, approximately 0.5. The structures with poorer matching between the target and predicted sets (i.e. $X_t$ and $X_p$ smaller than 0.25) are relatively far from the wall, and do not belong to the 95 % of the joint p.d.f.s – suggesting that there are very few of them.

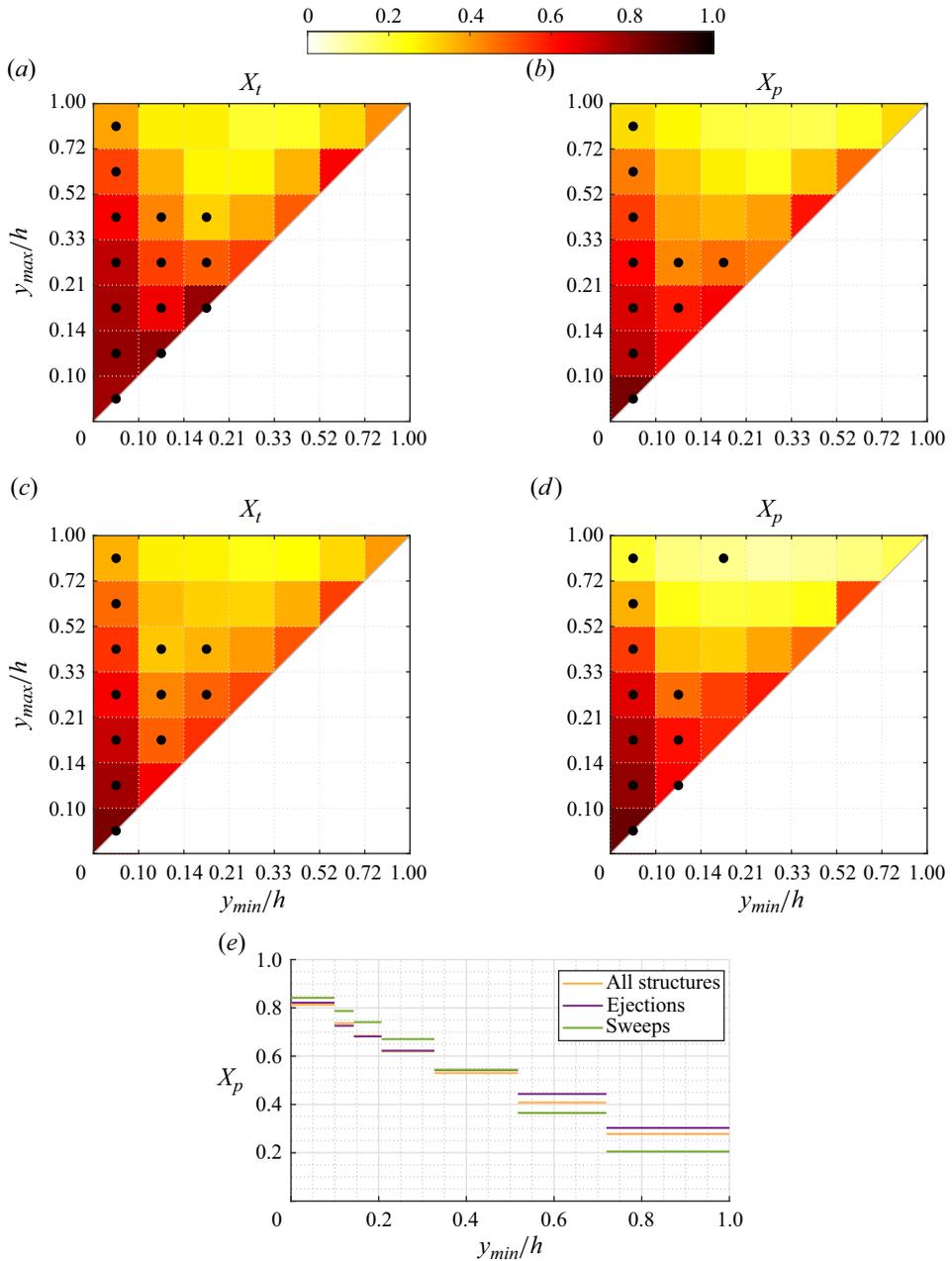

Figure 11. Maps of $X_t$ and $X_p$ for case A, considering only ejections-Q2 in $(a,b)$ and only sweeps-Q4 in $(c,d)$, analogous to those in figures 10$(a)$ and 10$(d)$ containing all structures together. Dotted bins represent the top 95 % of the joint p.d.f. of the structures over that target or prediction set, respectively. Profile $(e)$ of average matching proportion $X_p$ for case A, for the left column of bins ($y_{max}/h < 0.10$), comparing cases considering all the identified structures (figure 10$d$), only ejections (figure 11$b$) and only sweeps (figure 11$d$).

Several reasons may justify this performance: we expect a lower prediction ability from the 3-D GAN for wall-detached structures, for structures extending to higher $y_{max}/h$ and for types of structures that are not particularly common, with the computational resources available in the training process having been used to target other patterns within the flow.

The joint p.d.f.s for figure 10(*b*,*e*,*h*) and figure 10(*c*,*f*,*i*), which do not consider those regions in case A with few structures and poorer reconstructions, show smaller differences between the original and predicted sets of structures. In these cases, the top rows of bins in figure 10(*b*,*c*,*e*,*f*,*h*,*i*) are expected to include structures extending beyond its $y_{max}/h$ limit, cutting them and considering their respective reconstruction accuracy only up to its respective limit. As was also observed in figure 5 with the error trends, these metrics $X_t$ and $X_p$ also indicate a slightly improved prediction ability with the reductions in the volume of the domain considered.

With these distributions of average matching proportions $X_t$ and $X_p$ in each interval, a novel perspective on what the network is capable of reconstructing is given. The reconstructed volume of wall-attached structures is generally preserved (high $X_t$) and undistorted (high $X_p$) for wall-attached coherent structures. Even though the reconstruction precision in terms of volume and shape of the coherent structure seems to reduce progressively for increasing $y_{max}/h$, the values of $X_t$ and $X_p$ for wall-attached structures are still higher than for any other bin with $y_{min}/h > 0.1$ if $y_{max}/h < 0.5$. It can be argued that the reduction in reconstruction accuracy for increasing wall-normal distance is due prevalently to the progressively decreasing number of wall-attached structures, which should have an impact on the training of the 3-D GAN.

The increase in reconstruction fidelity when reducing the wall-normal thickness of the volume of the domain (i.e. cases B and C in figure 10) is in line with the hypothesis that the estimator focuses its effort in reconstructing features extending down to the wall. A marginal increase in $X_t$ is observed in bins corresponding to the same region for decreasing wall-normal thickness of the reconstructed volume. It can be hypothesized that the prevalence of wall-attached over detached – thus poorly correlated with flow quantities – structures in cases B and C simplifies the training of the network and improves its accuracy. Structures extending beyond the $y_{max}$ limit of each case are collected within the top row of bins of each plot in figure 10, and their reconstruction accuracy is slightly increased when the volume of the domain is reduced – although they are cut and a part of them is not being considered.

The $X_t$ and $X_p$ distributions share some similarities, with the ideas mentioned above. The main difference between them is the fact that the average matching proportions are slightly higher for $X_t$ than for $X_p$. From $X_t$, it is seen that with the behaviour just mentioned, the structures predicted by the network do not contain the whole volume of the original ones, denoting some loss of accuracy. Moreover, $X_p$ tells us that the reconstructed structures contain not only sections within the original structures but also regions out of them. With both metrics and their physical meaning, the combined effect of these two losses together is shown as $X_t X_p$ in figure 10. The superior reconstruction of these wall-attached structures must be highlighted, with a progressive loss with the wall-normal size $\Delta y$. The small structures lying right over the diagonal report a lower overall score than the attached ones, but higher than other wall-detached structures.

According to (3.1), turbulent structures are defined independently of the sign of $u(y)$ and $v(y)$, but imposing signs on them, the structures can be classified following the quadrant analysis (Lozano-Durán *et al.* 2012), with sweeps and ejections being of special interest. The maps shown in figure 11 allow us to compare and establish further conclusions and differences in the performance of the 3-D GAN when reconstructing sweeps and ejections. In all the cases, the left column of wall-attached structures is fully contained within the top 95 % of the joint p.d.f., with their $X_t$ or $X_p$ magnitude decreasing progressively with increasing $y_{max}/h$, as expected. In none of them is the family of wall-detached structures with small $\Delta y$ lying right above the diagonal a significant proportion of the population of structures. Hence these numerous structures highlighted in figure 10 may be mainly Q1 and Q3 structures. Moreover, at this threshold of the joint p.d.f., other structures that may be considered as attached although they are not in the leftmost column, are included in this top joint p.d.f. set. In terms of $y_{min}/h$, they extend from below 0.21 following this categorization in bins, while a very similar limit of 0.20 was set as the threshold to classify structures as attached or detached by (Lozano-Durán *et al.* 2012). However, in terms of $y_{max}/h$, the joint p.d.f. cut does not extend up to 1 – which is the case for $y_{min}/h < 0.1$ bins. For both sweeps and ejections, these numerous structures do not extend beyond the 0.52 limit for $X_t$ or beyond the 0.33 limit for $X_p$. This indicates that the network can reconstruct wall-attached sweeps and ejections with reasonable fidelity, although the part farther from the wall may be partially lost or even generated with poorer accuracy.

One of the main aspects reflected in figure 10 is the fact that more accurate reconstructions are reported for wall-attached structures, which are the most quantitative ones. The same is true for the distributions with sweeps and ejections as seen in figures 11(*a*)–11(*d*). For some of the bins right above the diagonal with wall-detached structures with small $\Delta y$, the quantities $X_t$ and $X_p$ reported are higher than for those bins of wall-attached structures with high $\Delta y$. Nevertheless, these wall-detached structures, as ejections (figures 11*a,b*) or sweeps (figures 11*c,d*), are not a big part of the population of structures, such that the training process may not focus substantially on them.

The distributions of $X_t$ and $X_p$ shown in figure 11 for ejections and sweeps are qualitatively similar, although some differences can be established. Considering $X_t$ for those bins within the region highlighted by the joint p.d.f., the metric is always higher for ejections than for sweeps, although they follow the same trends. The 3-D GAN can reconstruct wall-attached structures, generally and statistically preserving ejections slightly better than sweeps. One possible reason would be that the correlation with the wall measurements used is stronger with ejections, emerging from the wall, than with sweeps, which travel towards the wall. This could not be explained with the pressure measurements, which may be quite antisymmetric for both types of structures, according to studies assessing this correlation, such as the one by Sanmiguel Vila & Flores (2018). Nevertheless, this is not the case for the wall-shear stresses, also used as input data.

Regarding $X_p$, it is difficult to establish such a distinction between Q2 and Q4 structures. To compare these variations in $X_p$ along the structures with $y_{min}/h < 0.1$ more easily, figure 11(*e*) provides a different view. Wall-attached sweeps are less distorted than ejections and all structures overall when structures with short wall-normal heights are considered (up to $y_{max}/h < 0.3$). This trend is inverted for taller wall-attached sweeps ($y_{max}/h > 0.5$), reporting substantially lower $X_p$, with these sweeps being more distorted than ejections with similar $y$ size. Although the quantities are very similar and follow the same trends, the footprint impact of sweeps over the wall is stronger for the short structures close to the wall.

In addition to this, the change of trend experienced might also be explained from a statistical point of view. Ejections are lower in number than sweeps (see table 4), but occupy a much larger volume, while most sweeps and ejections are wall-attached. If most sweeps are attached but much smaller than ejections, then there may be less volume far from the wall occupied by wall-attached sweeps than by wall-attached ejections. There are some big sweeps extending from near the wall to the mid-plane, but they are not very frequent in the dataset, or not as much as big wall-attached ejections. Hence the 3-D GAN may learn better the patterns of those ejections, which are much more common. On the other side, close to the wall, wall-attached sweeps might be much more common than wall-attached ejections, so that the opposite happens.

## 4. Conclusions

A direct 3-D reconstruction from wall quantities with 3-D generative adversarial networks (GANs) has been proposed and demonstrated. The flow estimator builds on the successful reconstruction by Güemes *et al.* (2021) using a mapping from wall-shear stress and wall-pressure to 2-D wall-parallel velocity fields, and extends it to a full 3-D estimation. This extension comes with an affordable increase in the number of parameters and computational cost of the training if compared to the 2-D architecture estimating a single plane. The main advantage is a direct full reconstruction of the flow topology, without the need for training multiple networks for planar reconstruction. The argument for the reduction of the number of network parameters is the concept of parameter sharing, according to which the filters of the convolutional layers are shared among different wall-normal distances. Besides, the reconstruction based on this methodology, in which the main element of the network is the 3-D convolutional layer, ensures continuity within the reconstructed domains. In contrast, a procedure based on merging independently reconstructed 2-D planar domains could give rise to discontinuities.

The algorithm is tested on channel flow data at friction Reynolds number 200. The results in terms of reconstruction accuracy of the velocity fluctuations show a similar trend to the case of the 2-D single-plane estimators, with lower error on the streamwise velocity component with respect to the spanwise and wall-normal components. The error is in all cases slightly larger than in the reference case of the 2-D estimator with a similar Reynolds number. This was expected due to the comparably lower number of parameters per output node used in the 3-D estimator.

We also observe that a reduction in the target volume size does not always correspond to an improvement in accuracy. The estimator trained with test case D, which contained only the region with $y/h > 0.21$, performed worse in terms of reconstruction accuracy than the estimator of case A, whose target was the entire volume. This can be explained by the difficulty of the network to ignore the parts of the wall fields that were related only to structures located in $y/h < 0.21$ when trying to reconstruct the outer region of test case D. While in case A, a large portion of the network parameters is trained to establish the mapping between near-wall features and wall quantities, in case D, the estimator should learn to filter out the portion of wall quantities that is due to near-wall structures and at the same time is uncorrelated with the structures in the target volume. Due to the modulation effect of large scales on the near-wall cycle, there is an inevitable loss of accuracy in this process. While this might be frustrating in view of training neural networks that target the reconstruction of far-from-the-wall structures – which might be interesting for control

purposes – the higher performance of 2-D estimation with respect to case D hints at the possibility of accuracy improvement by increasing the number of parameters for this task.

The estimators for each of the cases, regardless of the reconstruction domain, seem to target specific features of the flow. In particular, wall-attached structures are reproduced with high fidelity at least up to $y/h = 0.5$ (i.e. 100 wall units), while a significant fraction of detached structures is filtered out in the process. This is a desirable feature since wall-attached structures carry the bulk of Reynolds stresses, and it was somewhat foreseeable due to their stronger wall footprint. We thus envision higher difficulty in predicting detached structures.

Furthermore, there are some differences in the prediction quality depending on the type of structure in relation to the footprint that they produce. Among wall-attached structures extending only in the near-wall region, sweeps are estimated slightly better than ejections. However, the opposite situation is found between wall-attached sweeps and ejections that extend up to the middle of the channel or close to it. The footprint of the structures to be reproduced is a key aspect in this process.

Sweeps and ejections dominate other types of structures. Although some of them, mainly wall-detached ones, are filtered by the GAN, the statistics of the identified structures in terms of volume and quantity are reasonably well preserved. There are more sweeps than ejections, but ejections are much bigger and occupy most of the volume identified as a structure according to the hyperbolic hole size employed.

**Supplementary material.** Data related to this work and trained models of the neural networks are openly available at https://doi.org/10.5281/zenodo.11090713. Codes developed in this work are openly available at https://github.com/erc-nextflow/3D-GAN.

**Funding.** This project has received funding from the European Research Council (ERC) under the European Union's Horizon 2020 research and innovation programme (grant agreement no. 949085, NEXTFLOW). Views and opinions expressed are, however, those of the authors only, and do not necessarily reflect those of the European Union or the ERC. Neither the European Union nor the granting authority can be held responsible for them. A.C.M. acknowledges financial support from the Spanish Ministry of Universities under the Formación de Profesorado Universitario (FPU) programme 2020. R.V. acknowledges financial support from ERC (grant agreement no. 2021-CoG-101043998, DEEPCONTROL).

**Declaration of interests.** The authors report no conflict of interest.

**Author ORCIDs.**

Antonio Cuéllar https://orcid.org/0000-0002-1023-6995;

Alejandro Güemes https://orcid.org/0000-0002-1673-9956;

Andrea Ianiro https://orcid.org/0000-0001-7342-4814;

Óscar Flores https://orcid.org/0000-0003-2365-0738;

Ricardo Vinuesa https://orcid.org/0000-0001-6570-5499;

Stefano Discetti https://orcid.org/0000-0001-9025-1505.

## Appendix A. Criteria for the design of the generator network

The networks proposed in this work have been trained on an NVIDIA RTX-3090 GPU. In addition to the necessary modifications to unlock mapping 2-D fields to 3-D domains, some changes to the network architecture were proposed in order to optimize it and obtain a more accurate flow estimation.

An important degree of freedom in defining the generator network is the position of the up-sampling layers, a tool commonly used for super-resolution purposes to make it possible to match the lower-resolution input with the higher-resolution output. Often, these layers are placed at the beginning, in the first layer or before the bulk of residual blocks (Dong *et al.* 2015; Wang *et al.* 2015), which would allow the following convolution layers to operate in a wider domain. Other authors prefer to define a gradual positioning of these layers (Osendorfer, Soyer & van der Smagt 2014). These two approaches, and in particular the first one, have the inconvenience that they produce models with 3-D convolutional layers much heavier than if the up-sampling layers were placed at the end of the network, which is another alternative (Shi *et al.* 2016). This latter option is found in the studies by Ledig *et al.* (2017) and Güemes *et al.* (2021). Concerning convolution layers, these become more complex and computationally demanding, depending not only on the size of the domain but also on the number of filters included. If the up-sampling layers are gradually placed with a moderate number of convolution filters, and a large number of filters is used only at the end for a few convolution operations, then the required computational resources can be maintained or even substantially reduced.

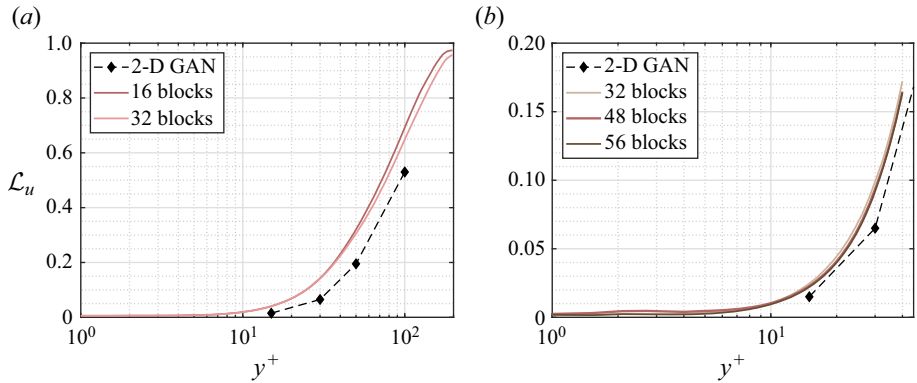

Figure 12. The MSE of the streamwise fluctuation velocity component $u$ for test cases (*a*) A and (*b*) C, with alternative network architectures.

Case A was first studied with a network comprising 16 residual blocks and all the up-sampling blocks after them. The convolutions in the residual blocks had 64 filters, and those in the up-sampling blocks had 256. As alternative architectures, several options have been tested, placing these up-sampling blocks not at the end, but right after specific residual blocks (as in figure 3 and table 2), which makes the convolution layers operate over broader domains in *y*. However, this change required convolution operations in up-sampling blocks to have 64 filters instead of 256, simplifying them and reducing the amount of trainable parameters. Indeed, this simplification is such that with the same machine and memory limitations, it allows us to increase the number of residual blocks further, from 16 to 32. This type of architecture is finally selected, as it reports a lower $\mathcal{L}$ error (2.1) even using less trainable parameters and requiring a comparable training time. This arrangement is a balance between different aspects, placing them progressively to allow convolutions to operate over wider domains than if they were at the end, while maintaining an efficient use of the computational resources. Both models can be compared in figure 12(*a*) and in table 5.

In view of the results from case A and their physical analysis and interpretation, cases B and C were proposed. Case B needs a special implementation, as the network needs to provide output data with 48 layers in the wall-normal direction, which is not a power of 2. Hence the first up-sampling layer increases this size from 1 to 3, and the subsequent ones continue as powers of 2 up to 48. For the rest of the set-up, the selected architectures for these two cases follow the same structure, with 32 residual blocks in total. Also, case

| Residual blocks | Filters per block | | Residual block up-sampling scheme | Trainable parameters | |
|---|---|---|---|---|---|
| | Residual | Up-sampling | | $G$ | $D$ |
| 16 | 64 | 256 | All at the end | $13.6 \times 10^6$ | $18.2 \times 10^6$ |
| 32 | 64 | 64 | 0-6-12-18-24-30 | $9.0 \times 10^6$ | $18.2 \times 10^6$ |

Table 5. Details on the implementation of alternative architectures for case A. The 'Residual block up-sampling scheme' column indicates the indexes of the residual blocks that are followed by an up-sampling block.

| Residual blocks | Filters per block | | Residual block up-sampling scheme | Trainable parameters | |
|---|---|---|---|---|---|
| | Residual | Up-sampling | | $G$ | $D$ |
| 32 | 64 | 64 | 6-12-18-24-30 | $8.0 \times 10^6$ | $23.8 \times 10^6$ |
| 48 | 64 | 64 | 8-16-24-32-40 | $11.6 \times 10^6$ | $23.8 \times 10^6$ |
| 56 | 64 | 64 | 10-20-30-40-50 | $13.3 \times 10^6$ | $23.8 \times 10^6$ |

Table 6. Details on the implementation of alternative architectures for case C. The 'Residual block up-sampling scheme' column indicates the indexes of the residual blocks that are followed by an up-sampling block.

D was proposed as the second half of the layers originally in case A not included in case C, concentrating the resources in this region of the domain. Although the reconstructed volume in case D is larger than in case C, both have 32 layers in the wall-normal direction (see table 1).

As for the discriminator, the same structure is followed in all the cases, as depicted in figure 4. Pairs of 3-D-convolution layers with increasing number of filters [64, 64, 128, 128, 256, 256, 512, 512] are used. From each pair, the first ones preserve the dimensions, and the second ones reduce the size of the data domain in the dimensions assigned to $x$, $y$ and $z$, the first time by a factor of 4, and thereafter by a factor of 2.

An additional change was proposed. The depth of the network can be modified easily by setting more or fewer residual blocks, such as the aforementioned change from 16 to 32 that showed an improvement in reconstruction terms. In this sensitivity analysis, case C was further trained with 48 and with 56 residual blocks, with up-sampling blocks gradually placed every 8 and every 10 residual blocks, respectively. These architectures and their performances can be compared in figure 12(*b*) and in table 6. They show a very moderate error reduction with respect to the previous situation with 32 blocks, with the curves practically overlapping each other, at the cost of using a substantially larger number of trainable parameters and taking longer to train. Without any remarkable improvements, these changes are discarded and the networks remain with 32 residual blocks for all cases.

## Appendix B. Comparison of the 3-D GAN with other methodologies

In § 1, GANs were set as the baseline for this work among other 2-D estimation techniques. Along the discussion of the results of this work, the accuracy of the proposed methodology is compared to that of 2-D GANs in terms of MSE. In this appendix, we provide a comparison of the performance of this network (refer to case A, from the wall to the mid-plane of the channel) with alternative techniques, i.e. LSE and EPOD as linear techniques, and a deep neural network (DNN) in the machine-learning framework. For the comparison with the 2-D GAN, it was convenient to compare the MSE of the velocity fluctuations normalized with their respective standard deviation at each wall-normal distance, as this normalization process is needed for both processes. For the comparisons that we are providing here, this normalization is not needed as the performances are obtained from the same dataset, and the results expressed here in inner units could be more meaningful and easier to analyse. As well as for the 3-D GAN, 16 000 samples are used to establish the correlation and 4000 to test the capabilities of each methodology.

### B.1. *Comparison of 3-D GAN and LSE*

One of the techniques that is used most often in the literature for flow estimation from wall measurements and other flow estimation purposes is LSE. The 3-D GAN convolutional filters act over multi-dimensional matrices, guaranteeing domain continuity. In LSE, the estimator works independently for the estimation of different points, and in particular, must be different for each wall-normal distance and for each velocity component to be estimated.

In this case, the number of sensors $n_s$ is 12 288, as three quantities are measured on a $64 \times 64$ grid. Each of the quantities $x'$ to be estimated can be computed through the projection of the vector $E[n_s \times 1]$ containing all the sensor measurements onto the vector $L[1 \times n_s]$ containing all the correlation coefficients:

$$x' = LE. \tag{B1}$$

Multiple samples are needed to compute the coefficients of $L$. To that end, multiple sensor entries are concatenated, defining a matrix where each column corresponds to a sample as follows:

$$E = [E_1|E_2|\cdots|E_t]. \tag{B2}$$

Given the known values of the quantities $x_i$ assigned to each set of sensor data $E_i$, the coefficients in $L$ are obtained from this linear system:

$$L(EE^{\mathrm{T}}) = [x_1, x_2, \ldots, x_t] E^{\mathrm{T}}. \tag{B3}$$

Once the coefficients that establish the correlation between the wall measurements and a velocity component at some point are known, the estimator (B1) can be used with new sensor inputs. It has been tested with 4000 samples. The output has been compared with the fields of the DNS to compare its performance with the 3-D GAN. The MSE of the three velocity fluctuations $[u^+, v^+, w^+]$ is shown in figure 13.

It is evident that the 3-D GAN outperforms the LSE estimator. The errors are very low close to the wall. The three velocity components experience a peak with a maximum error at $y^+ \approx 80$–90 with the 3-D GAN. With the LSE, these error peaks are larger and are shifted towards the wall, particularly for $u^+$, which has a substantially higher standard deviation (see figure 1). The errors of both techniques stabilize at approximately similar values near the mid-plane of the channel.

The use of the 3-D GAN has a clear benefit in the estimation of $u^+$, for which the highest errors are reported, with a maximum error almost three times lower with respect to the peak in LSE. The errors for the $v^+$ component, which has the lower standard deviation distribution along $y^+$, are significantly lower than for $u^+$, and the benefit between the 3-D GAN and the LSE is not so remarkable, although still important. The error of $w^+$ and its error reduction observed between the two methodologies are in an intermediate position between $u^+$ and $v^+$.

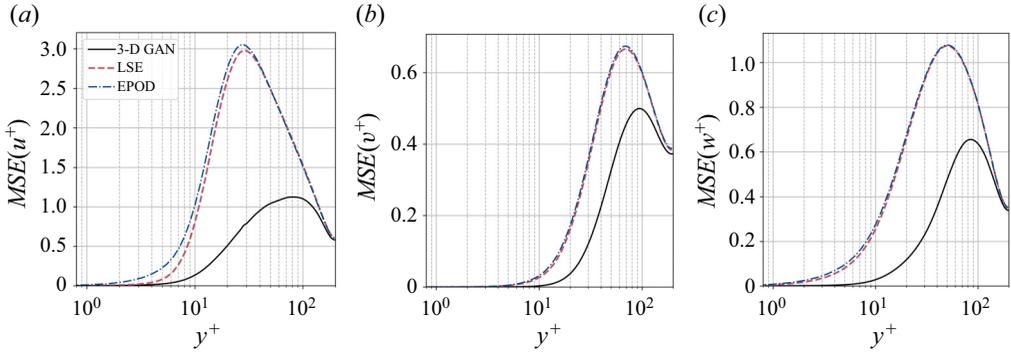

Figure 13. The MSE of the fluctuation velocity components (a) $u^+$, (b) $v^+$ and (c) $w^+$ of the 3-D GAN case A (solid line), the LSE (dashed line) and the EPOD (dot-dashed line). Results are given in wall-inner units.

Note that in the present analysis, we are not using the spectral formulation of LSE (SLES), which explicitly avoids spurious correlations between orthogonal Fourier modes (Encinar & Jiménez 2019). This choice is motivated by the interest in applying this technique to real-world applications, where the assumption of periodicity in the wall-parallel directions would be difficult to justify.

### B.2. *Comparison of 3-D GAN and EPOD*

The results obtained and discussed previously with the 3-D GAN are compared here with the performance of the EPOD estimator, whose methodology is described here.

The data of the velocity fluctuations, given in a multi-dimensional matrix, are rearranged in a 2-D matrix $X_U$, with one row for each sample (16 000 in this case), with the velocity components assigned to each of the points in space along the columns. The same procedure is done with the data from the wall probes, cast in the matrix $X_{pr}$. All this information is reduced in modes following the singular value decomposition, with the temporal information of each sample in $\Psi$ and the spatial information in $\Phi$, leading to

$$\left.\begin{array}{l} X_U = \Psi_U \Sigma_U \Phi_U^{\mathrm{T}}, \\ X_{pr} = \Psi_{pr} \Sigma_{pr} \Phi_{pr}^{\mathrm{T}}. \end{array}\right\} \tag{B4}$$

These matrices have been cropped, retaining the most energetic 7200 modes to remove those modes with low energy content that introduce noise in the problem. This threshold coincides with 99 % of the energy contained in $X_U$.

The final objective is to obtain the estimated (denoted with $*$) velocity fluctuations $X_U^*$ for a different set of samples, given those probe measurements. To that end, the temporal coefficients $\Psi_U^*$ associated with those samples are estimated and projected onto the spatial basis $\Sigma_U \Phi_U$ established previously as

$$X_U^* = \Psi_U^* \Sigma_U \Phi_U^{\mathrm{T}}. \tag{B5}$$

The temporal modes of the velocity field are obtained by projecting the temporal modes of the probes of the samples to be estimated onto the temporal correlation matrix $\Xi$ of the

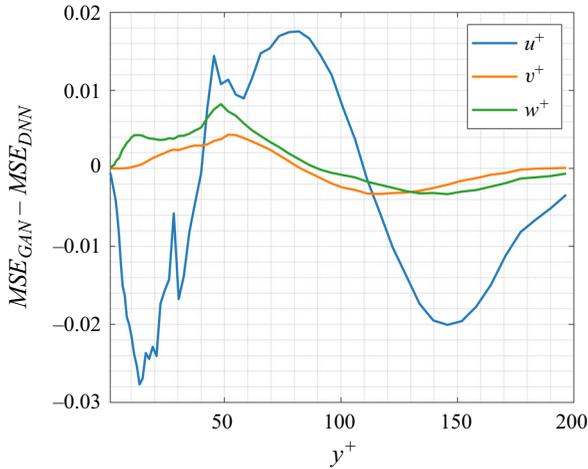

Figure 14. Difference of the MSEs of the fluctuation velocity components $u^+$, $v^+$ and $w^+$ of the 3-D GAN and the DNN. Results are given in wall-inner units.

probes and the velocity field:

$$\Psi_U^* = \Psi_{pr}^* \Xi = \Psi_{pr}^* \Psi_{pr}^{\mathrm{T}} \Psi_U. \tag{B6}$$

Again, the estimation of the 3-D GAN seems more faithful than that of the EPOD, as seen in figure 13. The behaviour of the error curve of the EPOD is very similar to that of the LSE, with the peak at a very similar $y^+$ and a small increment of error. The 3-D GAN, which incorporates nonlinearities in the problem, not only is capable of estimating the flow with a higher accuracy than these two linear techniques but also shifts the peak of maximum error away from the wall. This comparison shows the importance of performing the estimation with a nonlinear operator if accuracy farther from the wall is sought.

### B.3. *Comparison of 3-D GAN and DNN*

The performance of the 3-D GAN is also compared with a simpler concept of neural networks. To that end, this DNN replicates the generator network $G$ while it neglects the discriminator $D$. Its loss function is solely based on the MSE, as in (2.1), with zero contribution from the adversarial loss (2.2).

The difference in the error between these two methodologies is in general moderately low (figure 14). The main benefit of the 3-D GAN is observed in the estimation of the streamwise velocity fluctuations. Although in some regions ($y^+ \approx [40-110]$) the DNN estimates $u^+$ better than the 3-D GAN, the overall performance of the 3-D GAN is superior, with a clear advantage in terms of accuracy in the near-wall region and in the outer region, i.e. the most challenging one. The improvement is much less relevant for the estimation of $v^+$ and $w^+$. To provide a quantitative comparison of these results, the integrals of the area enclosed between these curves and the horizontal axis have been computed in the range $y^+ = 0-200$. The biggest benefit is found for $u^+$, reporting $-0.8027$. For $v^+$, there is a small benefit $-0.0131$, and for $w^+$ a small penalization $+0.1424$.